\RequirePackage{fix-cm}
\documentclass[smallextended]{svjour3}       
\smartqed  
\usepackage{natbib}

\usepackage{balance}
\usepackage{graphicx}
\usepackage{booktabs}
\usepackage{tikz}
\usepackage{multirow}
\usepackage{subfig}
\usepackage{tablefootnote}

\usepackage{listings}
\usepackage{array,booktabs,ragged2e}
\usepackage{soul}
\usepackage{marvosym}
\usepackage{textcomp}
\usepackage{tcolorbox} 
\newcommand{\rqbox}[1]{\begin{tcolorbox}[left=4pt,right=4pt,top=4pt,bottom=4pt,colback=gray!5,colframe=gray!40!black,before skip=6pt,after skip=6pt]#1\end{tcolorbox}}

\begin{document}

\title{An Exploratory Study on the Repeatedly Shared External Links on Stack Overflow}

\author{Jiakun Liu 
        \and Haoxiang Zhang
        \and Xin Xia \thanks{Xin Xia is the corresponding author.}
        \and David Lo 
        \and Ying Zou
        \and Ahmed~E.~Hassan
        \and Shanping Li
        }

\institute{Jiakun Liu \at
                College of Computer Science and Technology, Zhejiang University, China\\
                PengCheng Laboratory, China\\
         \email{jkliu@zju.edu.cn}
         \and
         Haoxiang Zhang \at
                 Queen's University, Canada\\
         \email{hzhang@cs.queensu.ca}
         \and
         Xin Xia \at
                Faculty of Information Technology, Monash University, Melbourne, Australia\\
         \email{xin.xia@monash.edu}
           \and
         David Lo \at
                School of Information Systems, Singapore Management University, Singapore\\
         \email{davidlo@smu.edu.sg}
         \and
         Ying~Zou \at
                 Queen's University, Canada\\
         \email{ying.zou@queensu.ca}
         \and
         Ahmed~E.~Hassan \at
                 Queen's University, Canada\\
         \email{ahmed@cs.queensu.ca}
         \and
         Shanping Li \at
         College of Computer Science and Technology, Zhejiang University, China\\
         \email{shan@zju.edu.cn}
        }

\date{Received: date / Accepted: date}

\maketitle

\begin{abstract}

On Stack Overflow, users reuse 11,926,354 external links to share the resources hosted outside the Stack Overflow website.
The external links connect to the existing programming-related knowledge and extend the crowdsourced knowledge on Stack Overflow.
Some of the external links, so-called as repeated external links, can be shared for multiple times.
We observe that 82.5\% of the link sharing activities (i.e., sharing links in any question, answer, or comment) on Stack Overflow share external resources, and 57.0\% of the occurrences of the external links are sharing the repeated external links.
However, it is still unclear what types of external resources are repeatedly shared.
To help users manage their knowledge, we wish to investigate the characteristics of the repeated external links in knowledge sharing on Stack Overflow.
In this paper, we analyze the repeated external links on Stack Overflow.
We observe that external links that point to the text resources (hosted in documentation websites, tutorial websites, etc.) are repeatedly shared the most.
We observe that: 1) different users repeatedly share the same knowledge in the form of repeated external links, thus increasing the maintenance effort of knowledge (e.g., update invalid links in multiple posts), 2) the same users can repeatedly share the external links for the purpose of promotion, and 3) external links can point to webpages with an overload of information that is difficult for users to retrieve relevant information.
Our findings provide insights to Stack Overflow moderators and researchers.
For example, we encourage Stack Overflow to centrally manage the commonly occurring knowledge in the form of repeated external links in order to better maintain the crowdsourced knowledge on Stack Overflow.

\end{abstract}

\maketitle

\section{Introduction}

Users share knowledge on Stack Overflow to solve programming related problems \citep{rahman2014towards, xia2017developers}.
For example, users are encouraged to share code when posting questions and answers by including code blocks, Stack Snippets\footnote{https://meta.stackoverflow.com/q/358992/}, and even the hyperlinks that point to the websites that host code resources, e.g., \textit{jsfiddle.net}.\footnote{https://stackoverflow.com/help/how-to-ask}$^, $\footnote{https://stackoverflow.com/help/minimal-reproducible-example}$^, $\footnote{https://stackoverflow.com/help/formatting}$^, $\footnote{https://stackoverflow.com/editing-help\#code}
However, when inspecting the shared code resources on Stack Overflow, previous work focused on extracting the code snippets from the post text bodies using the regular expression \lstinline{<pre><code>(.+?)</code></pre>} and overlooked the code resources shared by hyperlinks \citep{an2017stack, Chaiyong_2018_toxic, Wu2019, chen2015crowd, SO_code_reliable}.
Based on our analysis, users post 49,024,941 code snippets on Stack Overflow, while 4,198,150 external links also point to the code resources in code hosting websites.
Therefore, 1 out of 12 code resources is shared through external links.
Prior work never consider these external code resources in studying code on Stack Overflow.
This practice would introduce bias in the previous work.
For example, Chen et al. investigated the prevalence of the code related to the faulty implementation of security functionality on Stack Overflow by analyzing the shared code snippets on Stack Overflow \citep{SO_code_reliable}.
However, they did not consider all the code.
We observe that in an answer to how to secure service using passwords\footnote{https://stackoverflow.com/questions/28207373/}, answerers shared a link to a code snippet hosted in GitHub\footnote{https://github.com/jersey/jersey/blob/master/examples/https-clientserver-grizzly/src/main/java/org/glassfish/jersey/examples/h-ttpsclientservergrizzly/SecurityFilter.java} instead of posting the code snippet on Stack Overflow.
To improve the understanding of knowledge sharing on Stack Overflow, in this study we investigate what external resources are shared on Stack Overflow.

In question answering activities (i.e., posting a question, an answer, or a comment) on Stack Overflow, users are encouraged to share the existing programming-related knowledge with hyperlinks (i.e., clickable URLs in text blocks).\footnote{https://stackoverflow.com/editing-help}
Previous work focused on investigating the \textbf{internal link}s that share the resources hosted within the Stack Overflow website to analyze the evolution of the knowledge network that is connected by the internal links \citep{Ye2017}.
However, \textbf{external link}s are used more frequently to share the resources that are hosted outside the Stack Overflow website.
Based on our analysis of the Stack Overflow data that was released on Jun. 2, 2019, 20,082,160 posts and comments contain links, we observe that 82.5\% of the link sharing activities (i.e., sharing links in any question, answer, or comment) on Stack Overflow share external resources, while 21.3\% of these activities share internal links.
Although prior studies (e.g., \citep{8292806, baltes2020contextual}) analyzed documentation links on Stack Overflow, by far there is no study to analyze all the 11,926,354 external links on Stack Overflow.
In this paper, we are the first to mine all external links on Stack Overflow and investigate how external links contribute to the crowdsourced knowledge sharing activities on Stack Overflow.

Moreover, we observe that the external resources are not equally shared on Stack Overflow.
Some external links are shared for multiple times (i.e., \textbf{repeated external link}s), while others are shared for once.
For example, \lstinline{http://api.jquery.com/hover/} is an external link that documents the mouse event of hovering in JQuery.
This external link is shared for 1,258 times by 756 users on Stack Overflow.
Intuitively, an external link to be repeatedly shared can be associated with the link that can be beneficial to at least one programming-related problem.
However, it is still unclear what types of external links are repeatedly shared.
By characterizing the repeatedly sharing activities of different types of external links, we can understand which types of external links can be more beneficial to the software engineering process.

However, prior studies only characterize the links on Stack Overflow based on the resource types \citep{8292806, baltes2020contextual, Ye2017}.
To the best of our acknowledgment, we are the first study that characterizes the links on Stack Overflow from the aspect of frequencies.
More specifically, it is still unclear what are the characteristics of the repeated external links.
We would like to understand the risk of repeated external links when developers use Stack Overflow as a platform to seek answers to their programming-related problems.
For example, in a discussion related to the repeated external links on Stack Overflow\footnote{https://meta.stackoverflow.com/q/278115/}, the user would like to understand the risk brought by the repeatedly sharing of external links on Stack Overflow, e.g., spamming.
However, there is no exploratory analysis of the repeated external links on Stack Overflow.

Considering that repeated external links can be actively reused, we wish to understand what resources are repeatedly shared and what are the characteristics of the repeated external links during knowledge sharing on Stack Overflow.
To better characterize the external knowledge on Stack Overflow, we analyze all the 11,926,354 external links.
We first perform several preliminary studies of the external links on Stack Overflow.
We observe that the top 500 (i.e., 0.07\%) most shared external websites (i.e., set of pages and resources under the same domain) correspond to 14,633,033 (i.e., 71.5\%) occurrences of the external links.
We structure our study along with the following research questions:

\begin{enumerate}
\item \textbf{How prevalent are repeatedly shared external resources on Stack Overflow?}\\
17.0\% of the external links are repeated external links.
In 57\% of the occurrences of external links, the external links are repeatedly shared on Stack Overflow.
Answers with higher scores have a higher proportion of repeated external links.

\item \textbf{Which types of external websites are repeatedly shared on Stack Overflow?}\\
External links that point to text resources (i.e., textual resources that are hosted in documentation websites, tutorial websites, etc.) are the most repeatedly shared.
Image hosting websites and code hosting websites are the least repeatedly shared external websites to provide different examples for different questions.

\item \textbf{What are the characteristics of the external links that are repeatedly shared in Stack Overflow answers?}\\
Answers with more external links are associated with more revisions.
When the external links are repeatedly shared in different Stack Overflow answers, users are expected to repeatedly maintain the same external link (e.g., update invalid links in multiple posts), thus increasing the overall maintenance efforts.
12.3\% of the repeatedly shared external links in answers are from a single user for promotion purposes.

\end{enumerate}

Based on our findings, we provide actionable suggestions for Stack Overflow moderators and researchers.
For example, we encourage Stack Overflow to centrally manage the repeatedly shared links to alleviate the maintenance effort of external crowdsourced knowledge.
For future researchers, we suggest that they leverage the resources from external links, e.g., source code, when mining Stack Overflow.

\noindent\textbf{Paper Organization:} The remainder of the paper is organized as follows.
Section \ref{s_rw} describes the related work about link sharing in software engineering and the related work about mining Stack Overflow.
Section \ref{s_su} details our approach to collect and process the data that is used in our study.
Section \ref{s_observe} presents our findings of research questions.
Section \ref{s_discuss} discusses the implications of our study and provides actionable suggestions based on our findings.
Section \ref{threats} acknowledges key threats to the validity of our study.
Finally, Section \ref{conclusion} concludes our study and proposes potential future work.

\section{Related Work}\label{s_rw}
In this section, we discuss literature related to link sharing in software engineering and mining Stack Overflow.

\subsection{Link Sharing in Software Engineering}\label{ss_rw_knowledge_sharing}
One of the most related studies is the one by G'omez et al. \citep{gomez2013study}. They investigated link sharing on Stack Overflow by studying link types, website types, and the most shared links and websites. Compared to their work, our research concentrates on the repeated external links.

The work by Correa et al. and Wang et al. investigated the knowledge dissemination between the issue tracking systems and Stack Overflow \citep{correa2013integrating,wang2015automatic}.
More specifically, Correa et al. investigated the role and impact of Stack Overflow in issue tracking system \citep{correa2013integrating}.
They observed that the average number of comments posted in response to bug reports are less when Stack Overflow links are presented in bug report.
Wang et al. analyzed the links between the Android issues in bug tracking systems and Stack Overflow posts.
They proposed an automatic approach by integrating the semantic similarity between Android issues and Stack Overflow posts \citep{wang2015automatic}.
Baltes et al. \citep{baltes2020contextual} analyzed how and why documentation is referenced in Stack Overflow threads in two different domains
(Java regular expressions and Android development).
They found that links on Stack Overflow serve a wide range of distinct purposes, ranging from citation links attributing content copied into Stack Overflow, over links clarifying concepts using Wikipedia pages, to recommendations of software components and resources for background reading.
Different from their studies, our work characterizes all the external links (including documentation links) that are repeatedly shared on Stack Overflow.
By doing so, we could understand the link-sharing activity in terms of a specific behavior (i.e., repeatedly sharing), rather than the links related to a specific resource type.

Researchers also investigated the links between software engineering artifacts.
For example, Rath et al. investigated the inter-linking of commits and issues in open source projects and observed that among six large projects, 60\% of the commits are linked to issues \citep{rath2018traceability}.

On utilizing web resources, Xia et al. listed the frequency and difficulty of the different web search tasks performed by developers \citep{xia2017developers}.
Rahman et al. proposed a novel IDE-based web search solution that exploits three reliable web search engines (i.e.., Google, Bing, and Yahoo) and Stack Overflow through their API endpoints \citep{rahman2014towards}.

Similar to these studies, our work investigates link sharing activities in software engineering.
However, our research focuses on the repeated external links on Stack Overflow, rather than all the links, to understand the repeated sharing of different types of websites and to characterize the repeated external links during knowledge sharing on Stack Overflow.

\subsection{Mining Stack Overflow Knowledge}\label{ss_rw_mining_so}

Many researchers focused on characterizing the knowledge that is hosted on Stack Overflow.
For example, Barua et al. investigated the topic trends of Stack Overflow \citep{Barua2014}.
Rosen et al. studied what mobile developers ask about \citep{Rosen2016}.
Bajaj et al. mined Stack Overflow to discover the different types of questions asked by web developers \citep{Bajaj:2014:MQA:2597073.2597083}.
Linares-V'asquez et al. investigated the relationship between API changes in Android SDK and developers' reaction to those changes on Stack Overflow \citep{Linares-Vasquez:2014:ACT:2597008.2597155}.
Zhang et al. investigated the obsolete answers on Stack Overflow and observed that more than half of the obsolete answers were probably already obsolete when they were posted \citep{Zhang_2019}.
Saha et al. investigated why questions remain unanswered and concluded that the majority of them were due to low interest in the community \citep{Saha:2013:TUC:2491411.2494585}.

Researchers investigated the mechanisms of how users contribute their knowledge to Stack Overflow.
For example, Mamykina et al. observed that 1) the founders' close involvement with the community, 2) a highly responsive and iterative approach to design, and 3) a system of incentives that promotes desirable user behavior, contribute to the success of Stack Overflow \citep{Mamykina:2011:DLF:1978942.1979366}.
Cavusoglu et al. studied the badge system on Stack Overflow. Their findings confirmed the value of badges and the effectiveness of the gamification system \citep{Cavusoglu_badge}.

Researchers also investigated the collaborative editing activities on Stack Overflow.
For example, Li et al. observed that the benefits of collaborative editing on questions and answers outweigh its risks \citep{li2015is}.
Chen et al. observed that a large number of edits are about formatting, grammar and spelling \citep{chen_by_community}.
To assist users in making small sentence edits, they developed an edit-assistance tool for identifying minor textual issues in posts and for recommending sentence edits to make corrections.
They also developed a Convolutional Neural Network-based approach to learn editing patterns from historical post edits for identifying the need for editing a post \citep{chen_data_driven}.
Wang et al. analyzed whether revision-related badges have a negative impact on the quality of revisions \citep{wang2018users}.
They observed that 25\% of the users did not make any further revision once they received their first revision-related badge.

Considering the large amount of knowledge hosted on Stack Overflow, many researchers focused on the management of the crowdsourced knowledge.
For example, Xia et al. proposed a novel approach to recommend tags on Stack Exchange sites \citep{wang2014entagrec}.
Xu et al. designed a tool to recognize semantically relevant knowledge units on Stack Overflow \citep{Xu_Linkable}.
Anderson et al. designed a tool to determine which questions and answers are likely to have long-lasting value, and which ones are in need of additional help from the community \citep{Anderson:2012:DVC:2339530.2339665}.
Pal et al. investigated the evolution of experts in the Stack Overflow community and observed how expert users differ from ordinary users in terms of their contributions \citep{pal2012evolution}.
Hanrahan et al. developed indicators for difficult problems and experts, and examined how complex problems are handled and dispatched by multiple experts \citep{Hanrahan_modeling_difficulties}.

Many researchers investigated how to utilize the knowledge hosted on Stack Overflow to help with software engineering as well.
For example, Cai et al. and Huang et al. used the knowledge hosted on Stack Overflow to recommend APIs \citep{Cai_biker,Huang_biker}.
Chen et al. used the code blocks from Stack Overflow to detect defective code fragments in developers' source code \citep{chen2015crowd}.

Different from the these studies, our study analyzes the link sharing activities on Stack Overflow.
Users post 11,926,354 external links, thus bringing external knowledge into Stack Overflow. We wish to understand what external resources are actively shared by mining the external links.

\section{Case Study Setup}\label{s_su}
In this section, we present the dataset and the data processing steps that we used to extract the external links from the SOTorrent dataset.

\subsection{Dataset}\label{ss_su_dataset}

We perform our experiment with the SOTorrent dataset\footnote{https://zenodo.org/record/3255045\#.XYWaMyh3iUk} \citep{baltes2018sotorrent,baltes2019sotorrent}.
The SOTorrent dataset is based on the official Stack Overflow data dump from Jul. 31, 2008 to Jun. 2, 2019.

Baltes et al. extracted the links in all the versions of Stack Overflow posts and then store these links into table \texttt{PostVersionUrl} \citep{baltes2018sotorrent,baltes2019sotorrent}.
More specifically, they first extracted each version of the text and code from the post history that are stored in Stack Overflow data dump (i.e., table \texttt{PostHistory}).
Then they collected the links from the identified text with a regular expression.
One row in Table \texttt{PostVersionUrl} represents one link that was presented in the history of Stack Overflow posts.
Baltes et al. collected the Id of the posts that host the link, the PostHistoryId to indicate the versions that the link is presented in, and the PostBlockVersionId to indicates whether the link is posted in the text of the posts or the code blocks of the posts.
They also collected whether the link is shared in bare or in markdown format (i.e., column LinkType), the position of the link in the post (i.e., column LinkPosition), the anchor text of the link (i.e., column LinkAnchor), as well as the protocol, root domain, complete domain, path, query, fragment identifier, and the URL of the link.

Baltes et al. extracted the links in Stack Overflow comments with a regular expression and then store these links into table \texttt{CommentUrl}.
One row in Table \texttt{CommentUrl} represents one link that was presented in Stack Overflow comments.
Baltes et al. collected the Id of the comment that hosts the link, and the Id of the post that hosts the comment.
They also collected whether the link is shared in bare or in markdown format (i.e., column LinkType), the position of the link in the comment (i.e., column LinkPosition), the anchor text of the link (i.e., column LinkAnchor), as well as the protocol, root domain, complete domain, path, query, fragment identifier, and the URL of the link.
We encourage readers to read their work for the full details of the data collection process of the SOTorrent dataset.

\subsection{Link Extraction}\label{ss_su_link_extraction}

In our paper, we use the external links that are currently directly presented in the text of Stack Overflow posts (i.e., questions and answers) and comments as the experiment object.
Such external links are clickable and are used to share external resources from other websites.

To do so, we use the links that are presented in Table \texttt{PostVersionUrl}.
We use the \texttt{PostHistory} table from SOTorrent to extract the date when a post was created or modified to identify the links that are present in the latest version of Stack Overflow posts.
Finally, we obtain a list of links that are currently presented in Stack Overflow posts.

Zhang et al. observed that comments on Stack Overflow can be leveraged to improve the quality of their associated answers \citep{zhang2019comments}.
Therefore, we extract the links in comments from the \texttt{CommentUrl} table.
We also use the \texttt{Comments} table to retrieve the meta information of comments, e.g., the date when a comment was created, and the user who created a comment.
Finally, we obtain a list of links that are presented in Stack Overflow comments.

After we extract links from Stack Overflow posts and comments, we determine whether a link is an external link or an internal link using its root domain.
If the root domain of a link is \lstinline{stackoverflow.com}, the link is an internal link; otherwise, the link is an external link.
We do not consider the links with other domains under Stack Exchange\footnote{https://stackexchange.com/sites} as internal links because these websites cover a wide range of topics, such as culture, art, and business.
Finally, we obtain a list of external links and internal links that are currently presented in Stack Overflow posts and comments.

Table \ref{table_intermal_external_compare} present an overview of the extracted links.
It compares the external links with internal links on Stack Overflow in terms of the number of distinct links, the number of the Stack Overflow activities (i.e., posting a question, an answer, or a comment) that contain links, and the occurrences of links.
We observe that \textbf{the use of external links is more prevalent than that of internal links}.
Among the 14,905,705 distinct links on Stack Overflow, 11,926,354 (i.e., 80.0\%) of the distinct links are external links.
In contrast, only 2,965,764 (i.e., 20.0\%) of the distinct links are internal links.
Among the 20,082,160 posts and comments that contain links, 16,580,537 (i.e., 82.5\%) of the posts and comments have external links.
In contrast, only 4,208,002 (i.e., 21.3\%) of the posts and comments have internal links.
Therefore, external links are prevalently shared on Stack Overflow, and we wish to further investigate how the external links are repeatedly shared on Stack Overflow in Section \ref{s_observe_prevalence_repeat_link}.

  \begin{table*}[!ht]
    \centering
    \footnotesize
    \caption{
      Comparison of the prevalence of external links and internal links.
      The table highlights the popularity of external links over internal links. 
    }\label{table_intermal_external_compare}
\begin{tabular}{|m{0.11\linewidth}<{\raggedright}|m{0.1\linewidth}<{\raggedright}|m{0.1\linewidth}<{\raggedright}|m{0.11\linewidth}<{\raggedright}|m{0.11\linewidth}<{\raggedright}|m{0.12\linewidth}<{\raggedright}|m{0.13\linewidth}<{\raggedright}|}
    \hline
    \textbf{}      & \textbf{\# Distinct Links} & \textbf{\% Distinct Links} & \textbf{\# Posts \& Comments} & \textbf{\% Posts \& Comments} & \textbf{\# Occurrence} & \textbf{\% Occurrence} \\ \hline
    Internal Links & 2,979,351                  & 20.0\%                     & 4,279,551                         & 21.3\%                            & 4,762,734              & 17.1\%                 \\ \hline
    External Links & 11,926,354                 & 80.0\%                     & 16,580,537                        & 82.6\%                            & 23,040,432             & 82.9\%                 \\ \hline
    \end{tabular}
\end{table*}

\section{Results}\label{s_observe}

In this section, we present the results of our study.

\subsection{RQ1: How Prevalent are Repeatedly Shared External Resources on Stack Overflow?}\label{s_observe_prevalence_repeat_link}
\vspace{0.1cm}\noindent\textbf{Motivation:}
In Section \ref{ss_su_link_extraction}, we observe that external links are actively shared on Stack Overflow.
Among these links, the ones that are repeatedly shared can contribute significantly to the crowdsourced knowledge on Stack Overflow.
To understand how such repeatedly shared external links contribute to the crowdsourced knowledge on Stack Overflow, we study how many links are repeatedly shared and characterize these links.

\vspace{0.1cm}\noindent\textbf{Approach:}
To identify the repeated external links on Stack Overflow, for each external link that is extracted in Section \ref{ss_su_link_extraction}, we count its occurrences in any question answering activity.
\textbf{The occurrences of a link is the number of times that the link is shared for.}
We consider the repeated external links as the external links that are shared for over once on Stack Overflow.
Table \ref{citation_count} describes the distribution of the occurrences of the repeated external links.
We group the repeated external links based on the number of times of sharing each repeated external link.
For an interval of the number of times of sharing repeated external links, we present both the number and the occurrence count of links.

To describe the prevalence of repeated external links in Stack Overflow questions, answers, and comments, we count the occurrences of each external link in questions, answers, and comments, respectively.
We consider the repeated external links in different question answering activities as the external links that are shared for over once in that question answering activity.
Table \ref{table_repeat_link_post_type} describe the prevalence of repeated external links in Stack Overflow questions, answers, and comments, respectively.
We present the proportion of external links that are repeated external links, the proportion of the occurrences of external links that are repeated external links, and the average occurrences of repeated external links.

To characterize the correlation between repeated external links and the scores of the associated answers, we first extract the scores of the answers from SOTorrent Table \texttt{Posts}.
We observe that 2\% of the answers have the scores lower than 0 or higher than 100.
Therefore, we divide the answers into 12 phases based on their scores, i.e., a group of answers with the scores lower than 0, a group of answers with the scores higher than 100, and ten groups of answers with the scores that uniformly range from 0 to 100.
Then, we check the correlation between the proportion of the answers that share repeated external links among all the answers that share links and the scores of the associated answers.

\vspace{0.1cm}\noindent\textbf{Result:}
\textbf{17.0\% (i.e., 2,027,866) of the external links are repeated external links.
57.0\% (i.e., 13,141,944) of the occurrences of external links are the repeated external links.}
Each repeated external link is shared for 6.5 times on average and while the median is 2.0 times.
Table \ref{citation_count} lists the distribution of the occurrences of the repeated external links.
We observe that most of the repeated external links are shared for 2-10 times.
Only 316 repeated external links are shared for over 1,000 times (i.e., 615,204 times in total).
\lstinline{http://api.jquery.com/jQuery.ajax/} is the most repeatedly shared external link on Stack Overflow.
This link is shared 8,269 times in total.
Though a large collection of external resources are shared on Stack Overflow, only a small proportion of them are frequently reused.

\begin{table*}[!htb]
  \centering
  \footnotesize
  \caption{The distribution of repeated external links occurrences.
  Most of the repeated external links are shared for 2-10 times.
  }\label{citation_count}
  \begin{tabular}{|m{0.23\linewidth}|r|r|m{0.17\linewidth}|m{0.17\linewidth}|}
  \hline
  \textbf{\#Times of Sharing per Link} & \textbf{\#Links} & \textbf{\%Links} & \textbf{\#Occurrences} & \textbf{\%Occurrences} \\ \hline
  2-10                    & 1,816,761     &  89.6\%    & 5,556,684  &  42.3\%                 \\ \hline
  10-100                  & 199,818       &  9.9\%    & 4,613,865  &  35.1\%                 \\ \hline
  100-1,000               & 10,971        &  0.5\%     & 2,356,191  &  17.9\%                 \\ \hline
  \textgreater{}1,000     & 316           &  0.0001\%  & 615,204    &  4.7\%                  \\ \hline
  \end{tabular}
\end{table*}

\begin{table}[!htb]
  \centering
  \footnotesize
  \caption{The prevalence of the external links that are repeatedly shared in Stack Overflow questions, answers, and comments, respectively.
  External links in Stack Overflow answers are the most repeatedly shared.}\label{table_repeat_link_post_type}
\begin{tabular}{|l|m{3cm}<{\raggedright}|m{2cm}<{\raggedright}|m{1.8cm}<{\raggedright}|}
  \hline
  \textbf{Post Type} & \textbf{\% Repeated External Link Frequency Counts} & \textbf{\% Repeated External Link Counts} & \textbf{\# Average Occerrence Counts} \\ \hline
  Question           & 24.8\%                                               & 7.8\%                      & 3.9                             \\
  Answer             & 60.2\%                                               & 18.8\%                     & 6.5                             \\
  Comment            & 50.8\%                                               & 15.7\%                     & 5.5                             \\ \hline
  \end{tabular}
  \end{table}


\textbf{External links are more repeatedly shared in Stack Overflow answers than questions and comments}.
Table \ref{table_repeat_link_post_type} describes the external links that are repeatedly shared in Stack Overflow questions, answers, and comments, respectively.
18.8\% of the external links in Stack Overflow answers are repeatedly shared.
This proportion is 2.4 times and 1.2 times higher than the proportion in Stack Overflow questions and comments, respectively.
60.2\% of the occurrences of the external links in Stack Overflow answers are repeated external links.
This proportion is 2.4 times and 1.2 times higher than the proportion in Stack Overflow questions and comments, respectively.
The average times of repeatedly shared external links in Stack Overflow answers is 6.5.
This number is 1.7 times and 1.2 times higher than the number in questions and comments, respectively.

\begin{figure}[!htb]
\centering
\includegraphics[width = \linewidth]{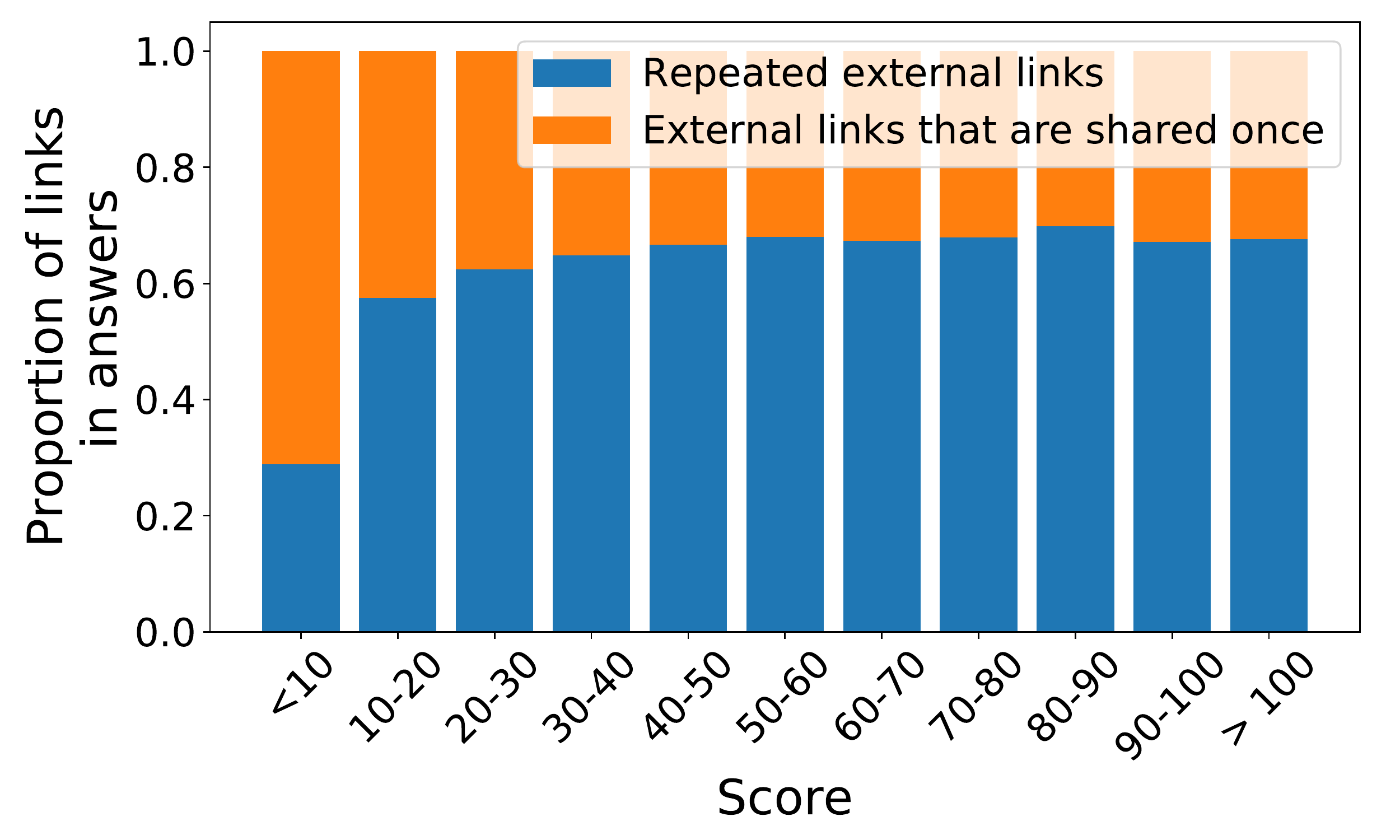}
\caption{The proportion of repeated external links among all links for answers with different score ranges.
The answers with higher scores have a higher proportion of repeated external links.}\label{figure_answer_score}
\end{figure}

\textbf{Answers with higher scores are associated with a higher proportion of repeated external links.
More specifically, among the answers and the accepted answers with external links, the proportions of repeated external links and the answer scores are significantly positively correlated}.
Figure \ref{figure_answer_score} shows the distribution of the proportions of repeated external links among all links for answers with different score ranges.
More specifically, we observe that the proportions of repeated external links and the answer scores are significantly correlated with Pearson's correlation coefficient = 0.674 (p-value $<$ 0.05) \citep{benesty2009pearson}.
For accepted answers, the proportions of repeated external links and the answer scores are significantly correlated with Pearson's correlation coefficient = 0.816 (p-value $<$ 0.05) \citep{benesty2009pearson}.

\rqbox{57.0\% of the occurrences of the external links are the repeated external links.
17.0\% (i.e., 2.0 million) of the external links on Stack Overflow are shared for multiple times, with a median occurrences number of 2.0.
Also, 18.8\% of the external links in Stack Overflow answers are repeatedly shared. This proportion is 2.4 times and 1.2 times higher than the proportion in Stack Overflow questions and comments, respectively.
Answers with higher scores have a higher proportion of repeated external links.
}

\subsection{RQ2: Which Types of External Websites are Repeatedly Shared on Stack Overflow?}\label{s_observeing_repeat_type}

\vspace{0.1cm}\noindent\textbf{Motivation:}
By using external links, users can introduce external resources outside the Stack Overflow website.
This practice extends the crowdsourced knowledge on Stack Overflow and enables users to leverage external resources.
However, previous work focused on the resources within the Stack Overflow website but overlooked the resources that are shared by external links.
In Section \ref{s_observe_prevalence_repeat_link}, we observe that external resources can be repeatedly shared in different questions, answers, and comments.
It is still unclear what external resources are repeatedly shared on Stack Overflow by external links.
To understand how the external links contribute to the crowdsourced knowledge on Stack Overflow, we analyze the types of the resources that are repeatedly shared by external links.

\vspace{0.1cm}\noindent\textbf{Approach:}
To analyze the types of the resources that are repeatedly shared by external links and what are the characteristics of these resources, we first study which external websites are the most commonly shared on Stack Overflow.
To identify the most commonly shared external websites, we first group the external links according to their root domains.
Then we count the occurrences of different websites and investigate the distribution of external websites.
As a result, we obtain of list of most shared websites.
The top 500 (i.e., 0.07\%) most shared websites host 14,633,033 (i.e., 71.5\%) of the external links on Stack Overflow.
The resources hosted on the top 500 most shared websites have high occurrences on Stack Overflow that introduce quantities of external knowledge to Stack Overflow.
We analyze the sharing of the external links that connected to the top 500 most shared external websites in this subsection.

To characterize the website types and the resource types of the top 500 most shared external websites, we perform two iterations of closed card sorting approach \citep{spencer2009card}.
We categorize these websites based on the definitions of website types (e.g., blog and official documentation) in Wikipedia.\footnote{https://en.wikipedia.org/wiki/Website}
The approach involves 2 iterations and is performed by the first two authors (i.e., A1-A2) in this paper:

\begin{itemize}
  \item Iteration I: A1 and A2 first randomly picked 100 websites from the top 500 most shared external websites.
  Then, A1 and A2 independently categorized these websites into different website types and resource types.
  More specifically, A1 and A2 randomly sampled 20 external links that are shared on Stack Overflow from each sampled website and browsed the sampled external links.
  A1 and A2 took notes regarding the deficiency or ambiguity of the labeling process for the sampled 100 websites.
  The inter-rater agreement of this iteration has a Cohen's kappa of 0.69, indicating that the agreement level is substantial \citep{viera2005understanding}.
  Next, A1 and A2 worked together to discuss the disagreements in the labeling process to resolve any disagreement.
  We observe that the website type is associated with the resource type.
  For example, the documentation websites host the text resources, and the image hosting websites host the visualization resources.
  During this phase, the coding schema of the website types and resource types of the external websites on Stack Overflow is revised and refined.
  
  \item Iteration II: A1 and A2 then independently labeled the remaining 400 websites of the top 500 most shared external websites.
  The overall Kappa value in this iteration is 0.76, indicating that the agreement level is substantial \citep{viera2005understanding}.
  This value is higher than that of the first iteration since the two labelers already had some experience to understand different website types and resource types.
  After completing the manual labeling process, A1-A2 and another post-doc worked together to discuss their disagreements to reach a common decision.
\end{itemize}

To depict the distribution of the occurrences of different types of websites, we traverse all external links that point to the top 500 most shared external websites.
Then in each website type, we count the number of websites, the number of distinct external links, and the occurrences of external links that point to the website.
Table \ref{table_website_type} describes the website types of the top 500 most shared external websites on Stack Overflow.
We also count the occurrences of the repeated external links, the number of the repeated external links, and the average times of the sharing of the repeated external links for each website type in the top 500 most shared websites.
Table \ref{table_repeated_domain} describes the statistics and characteristics of the repeated external links in different types of websites.

\begin{table*}[htb] 
  \centering
  \footnotesize
  \caption{The website types of the top 500 most shared external websites on Stack Overflow.
  We also highlight the resource types in bold.
  }\label{table_website_type}
  \resizebox{\linewidth}{!}{
  \begin{tabular}{|l|m{3cm}<{\raggedright}|m{2.4cm}<{\raggedright}|m{1.2cm}<{\raggedright}|m{1.2cm}<{\raggedright}|m{1.2cm}<{\raggedright}|}
  \hline
  \textbf{Type}  & \textbf{Function}                                        & \textbf{Example}  & \textbf{\% Websites} & \textbf{\% Links} & \textbf{\% Occurrences} \\ \hline
  Documentation          & Provide official documentation of a product                 & docs.oracle.com   & 16.4\%              & 9.0\%            & 18.2\%                      \\ \hline
  Official               & Provide a starting point to other resources of the product  & www.microsoft.com & 41.6\%              & 6.1\%            & 11.8\%                      \\ \hline
  Wiki                   & Provide readers with information with users' consensus      & en.wikipedia.org  & 2.0\%               & 0.9\%            & 3.0\%                       \\ \hline
  Blog                   & Post bloggers' personal ideas or personal documentation     & blogs.msdn.com    & 6.6\%               & 0.1\%            & 1.7\%                       \\ \hline
  Tutorial               & Provide online learning resources                           & www.w3schools.com & 3.6\%               & 0.4\%            & 1.2\%                       \\ \hline
  Forum                  & Provide message boards for opinion-based casual discussion  & forums.adobe.com  & 2.2\%               & 1.1\%            & 0.8\%                       \\ \hline
  Q\&A                   & Problem-solving websites                                    & askubuntu. com    & 3.0\%               & 0.7\%            & 0.9\%                       \\ \hline
  Bug tracker            & Bug-tracking, issue-tracking or project-management websites & issues.jboss.org  & 2.2\%               & 0.4\%            & 0.3\%                       \\ \hline
  \textbf{Text}          & \textbf{}                                                   & \textbf{}         & \textbf{77.6\%}     & \textbf{18.7\%}  & \textbf{37.9\%}             \\ \hline
  Image                  & Provide online image sharing and storage                    & i.stack.imgur.com & 1.6\%               & 22.1\%           & 11.7\%                      \\ \hline
  Video                  & Host and share videos                                       & youtube.com       & 0.8\%               & 0.5\%            & 0.4\%                       \\ \hline
  \textbf{Visualization} & \textbf{}                                                   & \textbf{}         & \textbf{2.4\%}      & \textbf{22.6\%}  & \textbf{12.1\%}             \\ \hline
  Code repository        & Share code projects                                         & github.com        & 2.0\%               & 9.9\%            & 9.4\%                       \\ \hline
  Runnable code          & Share runnable code examples                                & jsfiddle.net      & 6.0\%               & 14.3\%           & 8.2\%                       \\ \hline
  Code snippet           & Share code snippets that are not runnable                   & pastebin.com      & 0.4\%               & 1.6\%            & 0.9\%                       \\ \hline
  \textbf{Code}          & \textbf{}                                                   & \textbf{}         & \textbf{8.4\%}      & \textbf{25.8\%}  & \textbf{18.5\%}             \\ \hline
  File hosting               & Provide file hosting services                               & www.dropbox.com   & 6.8\%               & 1.8\%            & 1.9\%                       \\ \hline
  Invalid                & Currently unavailable                                       & java.net          & 2.8\%               & 0.6\%            & 0.4\%                       \\ \hline
  Search engine          & Response to a query with links                              & google.com        & 0.8\%               & 0.4\%            & 0.3\%                       \\ \hline
  URL shortener          & Provide URL shortener services                              & bit.ly            & 1.2\%               & 0.2\%            & 0.2\%                       \\ \hline
  \textbf{Other}         & \textbf{}                                                   & \textbf{}         & \textbf{11.6\%}     & \textbf{3.0\%}   & \textbf{2.8\%}              \\ \hline
  \end{tabular}
  }
\end{table*}

\begin{table*}\centering
  \footnotesize
  \caption{The statistics and characteristics of the repeated external links in different types of websites.
  We also highlight the statistics of resource types in bold.
  For example, 91.90\% of the occurrences of the links that point to the tutorial websites are repeated external links.
  47.30\% of the links that point to the tutorial websites are repeated external links.
  On average, the repeated external links that point to the tutorial websites are shared 12.6 times on average.
  }\label{table_repeated_domain}
  \resizebox{\linewidth}{!}{
  \begin{tabular}{|l|m{2cm}<{\raggedright}|m{1.5cm}<{\raggedright}|m{2cm}<{\raggedright}|m{6cm}<{\raggedright}|}
  \hline
  \textbf{Website Type} & \textbf{\% Occurrences of Repeated External Links} & \textbf{\% Repeated External Links} & \textbf{\# Average Occurrence Counts of Repeated External Links} & \textbf{Characteristics}                                                                                                                                                                           \\ \hline
  Tutorial              & 91.90\%                                            & 47.30\%                             & 12.6                                                             & Maintained by the public with consensus.                                                                                                                                                            \\ \hline
  Wiki                  & 90.70\%                                            & 42.30\%                             & 13.0                                                               & Maintained by the public with consensus.                                                                                                                                                            \\ \hline
  Documentation         & 83.50\%                                            & 36.10\%                             & 9.0                                                                & Provided by the official teams.                                                                                                                                                                     \\ \hline
  Official              & 82.10\%                                            & 33.70\%                             & 9.0                                                                & Provided by the official teams.                                                                                                                                                                     \\ \hline
  Blog                  & 80.50\%                                            & 34.20\%                             & 8.0                                                                & Provide personal viewpoints.                                                                                                                                                                        \\ \hline
  Q\&A                  & 66.60\%                                            & 16.20\%                             & 10.3                                                             & Provide guidelines, e.g., how to ask on Stack Overflow, which does not contribute to crowdsourced knowledge on Stack Overflow.                                                                                                                      \\ \hline
  Bug tracker           & 48.10\%                                            & 22.70\%                             & 3.1                                                              & Notify aksers of the identified bugs.                                           \\ \hline
  Forum                 & 46.00\%                                            & 17.80\%                             & 3.9                                                              & Provide detailed descriptions in a conversational way.                                                                                                                                              \\ \hline
  Text                  & 82.40\%                                            & 33.90\%                             & 9.1                                                              &                                                                                                                                                                                                    \\ \hline
  Code repository       & 56.50\%                                            & 20.40\%                             & 5.1                                                              & Provide code projects.                                                                                                                                                                              \\ \hline
  Code snippet          & 21.80\%                                            & 9.00\%                              & 2.8                                                              & Provide code snippets examples for different questions.                                                                                                                                             \\ \hline
  Runnable code         & 15.00\%                                            & 6.00\%                              & 2.8                                                              & Provide runnable code examples for different questions.                                                                                                                                             \\ \hline
  Code                  & 36.30\%                                            & 11.70\%                             & 4.3                                                              &                                                                                                                                                                                                    \\ \hline
  Video                 & 38.00\%                                            & 15.00\%                             & 3.5                                                              & Provide lengthy content and is lack support to search \citep{Ponzanelli_2016_video}, \citep{macleod2015code}, \citep{Ponzanelli_2016_codetube}. \\ \hline
  Image                 & 4.00\%                                             & 1.80\%                              & 2.2                                                              & Provide visualization example for different questions.                                                                                                                                              \\ \hline
  Visualization         & 5.20\%                                             & 2.20\%                              & 2.5                                                              &                                                                                                                                                                                                    \\ \hline
  Search engine         & 38.50\%                                            & 6.30\%                              & 9.2                                                              & Users use different keywords to search for different things.                                                                                                                         \\ \hline
  Invalid               & 36.70\%                                            & 12.90\%                             & 4.0                                                                & Cannot provide long-lasting value to viewer.                                                                                                                                                                 \\ \hline
  URL shortener         & 44.50\%                                            & 7.00\%                              & 10.8                                                             & The URL shortener websites make different users share the same resource with different links.                                                        \\ \hline
  File hosting              & 61.80\%                                            & 22.10\%                             & 5.7                                                              & Provide resources for users to download.                                                                                                                                                            \\ \hline
  Other                 & 54.60\%                                            & 17.20\%                             & 5.8                                                              &                                                                                                                                                                                                    \\ \hline
  \end{tabular}
  }
\end{table*}

\begin{figure}[!htb]
  \centering
  \includegraphics[width = \linewidth]{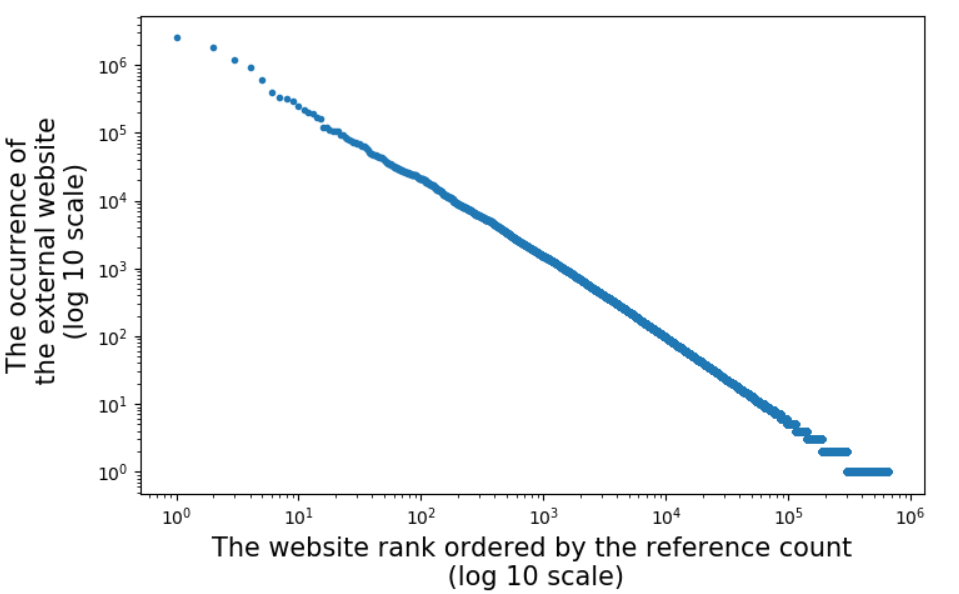}
  \caption{The occurrences of the external websites on Stack Overflow}\label{power_law}
\end{figure}

\vspace{0.1cm}\noindent\textbf{Results:}
\textbf{The top 500 (i.e., 0.07\%) most shared websites host 14,633,033 (i.e., 71.5\%) of the external links on Stack Overflow.}
There are 661,953 external websites on Stack Overflow.
Fig. \ref{power_law} shows that the occurrences of different external websites conform to the power-law distribution with $\alpha = 1.8$ and $xmin = 5.0$ \citep{power_law}.
For example, the image hosting external website, \lstinline{i.stack.imgur.com}, is the most shared website (i.e., 2,527,767 times in total).
Stack Overflow uses the image hosting service provided by \textit{i.stack.imgur.com} to mitigate the bandwidth and the storage burden of Stack Overflow\footnote{https://meta.stackexchange.com/q/90342}.\footnote{We consider the links to \lstinline{i.stack.imgur.com} as external links because 1) the resources are hosted in different domains, 2) the resources hosted on \lstinline{i.stack.imgur.com} is not maintained by Stack Overflow. It can be the case that the posts that share the links to \lstinline{i.stack.imgur.com} exist but the links are broken, which is similar to other external links.}
\lstinline{github.com} is the second most shared website (i.e., 1,870,707 times) on Stack Overflow.
And \lstinline{jsfiddle.net} is the third most shared website (i.e., for 1,187,120 times) on Stack Overflow.
It enables users to test their JavaScript, CSS, HTML or CoffeeScript code online and obtain the view after rendering.

\textbf{Text resources, such as the resources hosted in documentation websites and official websites, are the most commonly shared on Stack Overflow as 37.9\% of the occurrences of external links are sharing the text resources}.
During our labeling process, we observe that different types of websites offer the same type of resources.
Table \ref{table_website_type} shows our labeling result on the types of websites.
We observe that 77.6\% of the top 500 most shared external websites are related to text resources.
Text resources can come from multiple sources, such as documentation websites, tutorial websites, and bug tracker websites.
Moreover, 37.9\% of the occurrences of the external resources are related to the text resources.
For example, the website, \lstinline{msdn.microsoft.com}, is a technical documentation site that is shared the most (i.e., for 868,552 times) on Stack Overflow.
The website hosts the documentation related to the technologies developed by Microsoft.

Stack Overflow encourage users to share code in posts so that readers can better understand the problems.\footnote{https://stackoverflow.com/help/minimal-reproducible-example}
Stack Overflow introduce code blocks\footnote{https://stackoverflow.com/editing-help\#code}, and Stack Snippets\footnote{https://meta.stackoverflow.com/questions/358992/ive-been-told-to-create-a-runnable-example-with-stack-snippets-how-do-i-do} for users to paste code snippets.
By analyzing the whole Stack Overflow data dump, we observe that code resources also can be commonly shared the links that point to code hosting websites.
More specifically, users post 49,024,941 code snippets on Stack Overflow, while 4,198,150 external links also point to the code hosting websites.
This indicates that \textbf{53,223,091 code resources are shared with 7.9\% of them are shared through external links.}
We suggest that researchers should leverage the external code resources when mining Stack Overflow.

\textbf{External links that point to the text resources are the most repeatedly shared, as 82.4\% of the occurrences of the text resources are the repeated external links}.
Table \ref{table_repeated_domain} describes the repeated sharing of different types of websites.
More specifically, 82.4\% of the occurrences of the text resources are the repeated external links.
This proportion is 15.8 times higher than the proportion of the other types of resources.
33.9\% of the external links that point to the text resources are repeated external links.
This proportion is 15.4 times higher than the proportion of other types of resources.
On average, each repeated external links that points to text resources is repeatedly shared for 9.1 times.
This value is 3.6 times higher than the value of other types of resources.
The external links that point to the text resources have a higher impact compared with the external links that are associated with other types of resources.
We encourage researchers to investigate the repeatedly shared text resources on Stack Overflow.

For example, among all types of websites on Stack Overflow, tutorial websites are the most repeatedly shared external websites, as 91.9\% of the occurrences of the tutorial resources are the repeated external links.
More specifically, 47.3\% of the external links that point to the tutorial websites are repeated external links.
On average, each repeated external link that points to tutorial websites are associated with 12.6 questions on Stack Overflow.
For example, \lstinline{http://www.w3schools.com/sql/sql_join.asp} is widely used in 673 question threads as a walk-through tutorial to learn how to combine rows from two or more tables based on a related column between them.

\textbf{We observe that 32.3\% of the external links to bug tracker websites are posted in accepted answers.
This proportion is 1.3 times higher than the proportion of all external links that are ever posted in accepted answers.}
One possible reason is that the askers are not aware of the identified bugs before asking on Stack Overflow.
Below is an example\footnote{https://stackoverflow.com/q/28886508/} of an external link to a bug tracker website\footnote{http://bugs.python.org/issue22942}:

\vspace{0.1cm} \hangindent 1.5em\textit{   It appears to be a documentation bug for which \underline{an issue} has been raised.
}\vspace{0.1cm}

\noindent\normalsize In this example, the answerer responsed with an external link that points to a bug tracker website.
The asker commented that \textit{Thank you for the help. I didn't know about the issue tracker!}
The asker acknowledged the identified bugs by the external links that point to the bug tracker websites in Stack Overflow answers.
In the future, Stack Overflow can leverage the existing issues from bug tracking systems.
When a user is in the progress of writing a question, Stack Overflow could display a summary of possible related bugs.

\textbf{External links that point to image hosting websites and code hosting websites are the least repeatedly shared, as the average occurrences of the repeated external links in these types of websites are the smallest (i.e., less than 3)}.
We observe that external links to visualization resources are the least repeatedly shared external resources, as only 1.8\% of the external links that point to image websites are repeatedly shared.
4.0\% of the occurrences of the image websites are sharing the repeated external links.
One possible reason is that the images are used to provide visualization examples, and different questions are expected to use visualization examples case by case.

\textbf{External links that point to search engine websites in Stack Overflow answers provide evidence that users may not know how to search properly}.
52,331 external links to search engine websites are prevalently shared but are less repeatedly shared on Stack Overflow, as only 6.3\% of the external links that point to such websites are repeatedly shared.
One possible reason is that users use different keywords to search for different things.
Below is an example\footnote{https://stackoverflow.com/q/22021491/} use of an external link to a search engine website.
Stack Overflow encourage askers to ``search and research'' before posting questions.\footnote{https://stackoverflow.com/help/how-to-ask}
In the question, the asker shows his visited ``search and research'' link that points to a search engine website, i.e., google.com.
But this ``search and research'' link cannot solve his problem:

\vspace{0.2cm}\hangindent 1.5em\textit{The Linux cross reference services show no usage of them. \underline{Googling}\footnote{https://www.google.com/\#q=\_IOWR\_BAD+OR+\_IOR\_BAD+OR+\_-IOW\_BAD\&safe=off} turns up about a dozen Linux kernel cross reference sites, but absolutely no usage.
}\vspace{0.2cm}

\noindent\normalsize The accepted answer\footnote{https://stackoverflow.com/q/22021491/22021641} shows the explanation to his problem and suggests that:

\vspace{0.2cm}\hangindent 1.5em\textit{And in general, the search terms you are probably looking for is \underline{\_IOR\_BAD lkml}\footnote{https://www.google.com/search?q=\_IOR\_BAD+lkml}
}\vspace{0.2cm}

\noindent\normalsize In fact, a slight change to the keywords of a query can make a difference in the quality of search results.
Users make a nontrivial effort in searching and still end up asking a question.
In contrast, an expert can solve the question with a better search query.
In the future, researchers could design a tool to help Stack Overflow users better search on search engine.
A possible recommendation system of how to search the crowdsourced knowledge can be designed by leveraging the query patterns of expert users on Stack Overflow.

\textbf{The accessibility of external links to URL shortener websites not only depends on the accessibility of the target links but also depends on the stability of the URL shortener service}.
On Stack Overflow, the maximum length of a post is 30,000 characters\footnote{https://meta.stackexchange.com/q/176445}.
External links that point to URL shortener websites are prevalently shared but are less repeatedly shared on Stack Overflow, as only 7.0\% of the external links that point to URL shortener websites are repeatedly shared.
One possible reason is that the URL shortener websites make different users share the same resource with different links.
Below is an example\footnote{https://stackoverflow.com/q/2660914} of an external link to a URL shortener website\footnote{http://goo.gl/b93ns}:

\vspace{0.1cm} \hangindent 1.5em\textit{   take a look at the source code of this page, may be it will get you going in the right direction...
\underline{http://goo.gl/b93ns}
}\vspace{0.1cm}

\noindent\normalsize The asker asks about the YouTube player API \textit{addEventListener()}.
The answerer responses with an external link that points to a URL shortener website.
However, the external link is dead and we cannot get valid information from this answer.
This example shows that the accessibility of the external links that point to URL shorteners depends on the accessibility of the target links and the stability of the URL shortener service.
The URL shorteners can obfuscate the links lead and ultimately make the answer less useful.
When the URL shortener service is down, there is no way for users to recover the target location of the links.
Therefore, we recommend Stack Overflow could replace the external links to URL shortener websites with the target links in full length.

\rqbox{Text resources are the most repeatedly shared on Stack Overflow.
External links to image hosting websites and code hosting websites are the least repeatedly shared to provide different examples for different questions.
}

\subsection{RQ3: What are the Characteristics of the External Links that are Repeatedly Shared in Stack Overflow Answers?}\label{s_observeing_post}

\vspace{0.1cm}\noindent\textbf{Motivation:}
Stack Overflow users share their knowledge in Stack Overflow answers to solve programming-related problems.
By browsing the Stack Overflow answer, askers and viewers can obtain the solution of the questions.
In Section \ref{s_observe_prevalence_repeat_link}, we observe the repeated external links are the most prevalent in Stack Overflow answers, compared with questions and comments.
However, it is still unclear how the external links are repeatedly shared in Stack Overflow answers (we refer to the repeated external links as the external links that are repeatedly shared in Stack Overflow answers in this section).
To characterize how the repeated external links contribute to the problem-solving activities on Stack Overflow, we study the repeated external links in answers from different perspectives, e.g., the maintenance effort, the users who share the repeated external links, and the content of the repeated external links.
To better understand the organization of the knowledge that is connected by the same repeated external links, we also analyze the relations of the answers that share the same repeated external links.

\vspace{0.1cm}\noindent\textbf{Approach:}
We conduct both a quantitative study and a qualitative study to characterize the external links that are repeatedly shared in Stack Overflow answers.

In the quantitative analysis, we captured an overall picture of the repeated external links.
Users can conveniently introduce external knowledge to the crowdsourced knowledge on Stack Overflow by external links.
Some of the external links can be repeatedly shared on Stack Overflow to repeatedly share the same external resources.
We first analyze whether the use of repeated external links can alleviate the maintenance effort of the Stack Overflow community.
We scan the PostVersionUrl Table and compare the consecutive two versions, to track whether a newer version of the post revised the external links, e.g., removed the link, updated the text anchor of the link in the post.
Then, we count the number of revisions that are conducted to each external link in history.
This number is the maintenance effort that the Stack Overflow community has paid to maintain each external link.
Finally, we count the number of revisions related to the external links that are posted in the latest versions of Stack Overflow posts.
For the external links in the latest version of Stack Overflow posts, we can tell which external links were repeatedly revised in history.
The external links in the latest version of Stack Overflow posts that are with more revisions in history have the threat to be associated with more revisions in the future.

To obtain an overview of the repeated external links that are shared by different users, for each repeated external link, we calculate the normalized entropy of the numbers of times that different users share that external link in Stack Overflow answers (shortened as user entropy).
Besides, we are interested in whether the repeated external links with different user entropies are related to the quality of the repeated external links.
To do so, we count the number of the repeated external links that are ever posted in accepted answers in different users entropies.

We also inspect the repeated external links from the perspective of the content of the external links.
Intuitively, external links that are shared with the anchor part (i.e., the links with a hash mark \# in the last part, shortened as external link$_{anchor}$), can identify a portion of the resources that are shared in the bare links (i.e., the links with the anchor part that are shared without the anchor part, shortened as external links$_{no\_anchor}$) \citep{berners2005rfc}.
However, an external link$_{no\_anchor}$ does not illustrate which part of the link is related to the Stack Overflow answers.
We refer such external links$_{no\_anchor}$ as information overloading external links$_{no\_anchor}$.
For example, the external link$_{anchor}$, \lstinline{https://git-scm.com/docs/git-rebase#_merge_strategies}, could locate the resource hosted on the webpage that is pointed by the external link$_{no\_anchor}$, \lstinline{https://git-scm.com/docs/git}-\lstinline{rebase}.
When browsing this external link$_{no\_anchor}$, users would observe too much information that describes the git rebase from different perspectives.
We use the prevalence of the sharing of external links$_{no\_anchor}$ as an indicator to investigate the prevalence of sharing of the information overloading repeated external links.

\begin{table}[!htb]
  \centering
  \footnotesize
  \caption{Reasons for the Answers Pairs that Share the Same Repeated External Links}\label{table_answer_reason}
  \begin{tabular}{|m{0.2\linewidth}<{\raggedright}|m{0.5\linewidth}<{\raggedright}|m{0.05\linewidth}<{\raggedright}|}
  \hline
  \textbf{Reason}          & \textbf{Definition}                                                                     & \textbf{\%} \\ \hline
  Repeated Knowledge & Different users repeatedly share the same knowledge in the form of repeated external links. & 75\%        \\ \hline
  User Spam                & The same users for promotion purposes.                                                   & 15\%        \\ \hline
  Information Overloading     & The content of the external links is information overloading.                              & 10\%        \\ \hline
  \end{tabular}
\end{table}

Then in the qualitative analysis, we study the relations for the answers that share the same repeated external links in Stack Overflow answers.
To do so, we need a list of answers pairs that contain the same repeated external links.
More specifically, we first identify the repeated external links that are repeatedly shared in Stack Overflow answers.
Then, for each repeated external link, we list the answers that share the link.
We store all the combinations of the answers as the list of answers pairs that contain the same repeated external links.
For each type of resource, we randomly sample 100 answers pairs with a 95\% confidence level and 10 confidence interval.
These 300 answers pairs are our manual classification objects.
Note that a pair of answers can have multiple relations for sharing the same repeated external links.
To analyze the relationship of different answers that share the same repeated external link, the first two authors (i.e., A1-A2) perform two iterations of the card sorting approach \citep{spencer2009card}:

\begin{itemize}
    \item In the first iteration of labeling, we randomly sampled 50 answer pairs from the 300 sampled answer pairs.
    Then, A1 and A2 independently categorized the types of relationships and took notes regarding the deficiency or ambiguity of the labeling for the answer pairs.
    The inter-rater agreement of this iteration has a Cohen's kappa of 0.76, indicating that the agreement level is substantial \citep{viera2005understanding}.
    Next, A1 and A2 worked together to discuss the disagreements in the labeling process to resolve any disagreements until a consensus was reached.
    During this phase, we determined the types of relations, as is shown in Table \ref{table_answer_reason}.
    \item In the second iteration of labeling, A1 and A2 then independently labeled the remaining of 250 sampled answer pairs using the aforementioned approach.
    The overall Kappa value in this iteration is 0.79, indicating that the agreement level is substantial \citep{viera2005understanding}.
    This value is higher than that of the first iteration since the two labelers already had some experience to understand different relations for sharing the same repeated external links.
    After completing the manual labeling process, the two labelers and another post-doc worked together to discuss their disagreements to reach a common decision.
\end{itemize}

\vspace{0.1cm}\noindent\textbf{Result:}
\subsubsection{Quantitative Results}

\begin{figure}[!htb]
  \centering
  \includegraphics[width = \linewidth]{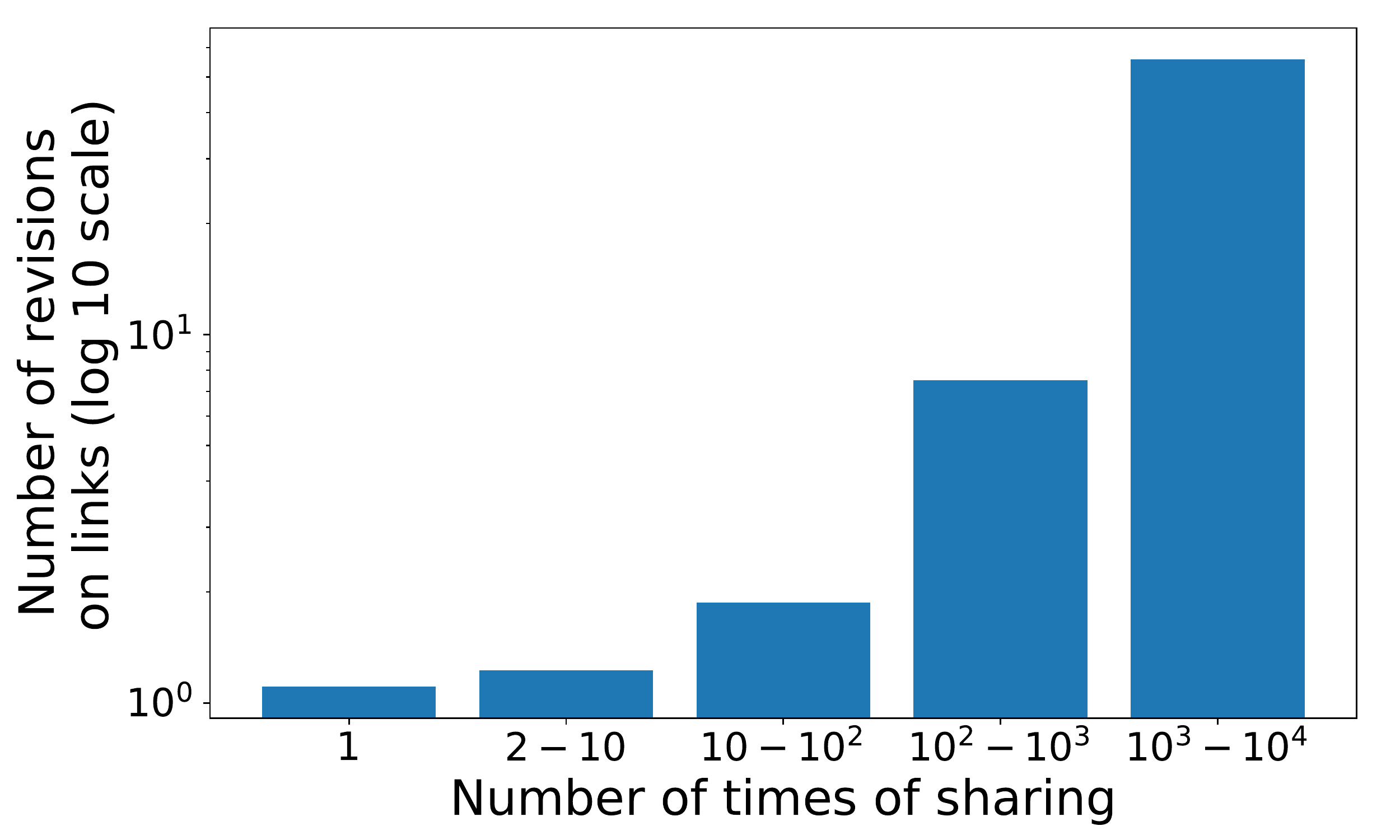}
  \caption{The distribution of the number of answers that revised the repeated external links based on the number of times of sharing repeated external links in answers.}\label{figure_mod}
\end{figure}

\textbf{External links that are more repeatedly shared are associated with more revisions}.
Figure \ref{figure_mod} shows the distribution of the number of versions that revise the external links based on the number of times of sharing external links in answers.
More specifically, we observe that the number of versions that revise the external links and the number of times of sharing external links are significantly correlated with Pearson's correlation coefficient = 0.7374 (p-value $<$ 0.05) \citep{benesty2009pearson}.
This shows that the use of repeated external links increases the maintenance effort.
Once the content of the external links is updated, the answers that involve the external links are expected to be updated.
Besides, once the content of the link is invalid, the Stack Overflow users have to replace the link with another qualified link.
Stack Overflow users have to repeatedly maintain the same repeated external links for multiple times.
This practice would increase the overall maintenance effort of the Stack Overflow.
We encourage the Stack Overflow could centrally maintain the external links.
Users can share the resources that are hosted and maintained by the Stack Overflow.
This practice can decrease the dependencies on external websites.

\begin{figure}[!htb]
  \centering
  \includegraphics[width = \linewidth]{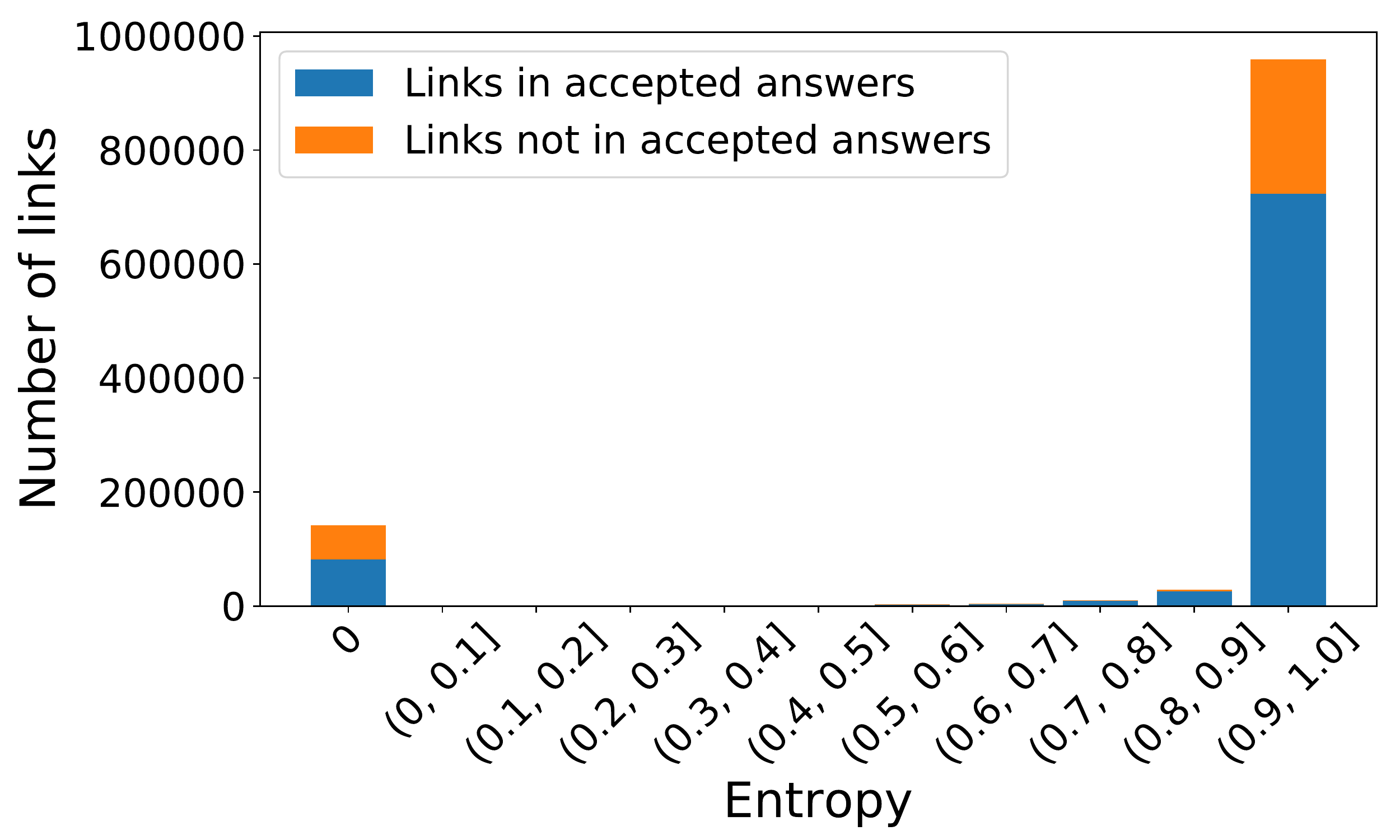}
  \caption{The distribution of the numbers of links based on the normalized entropy of the number of times that different users share the repeated external link.
  Most of the repeated external links are shared by different Stack Overflow users, while 13.6\% of the repeated external links are only shared by one user.}\label{figure_entropy}
\end{figure}

\textbf{83.4\% (i.e., 959,253) of the repeated external links are shared uniformly by different users, indicating that the value of such links is appreciated by the whole Stack Overflow community.
In contrast, 12.3\% (i.e., 141,561) of the repeated external links are shared by one user.
The proportion of the repeated external links that are shared by one user in accepted answers is less than the proportion of other repeated external links.}
Figure \ref{figure_entropy} plots the distribution of the numbers of the repeated external links based on the normalized entropy of the number of times that different users share the specific repeated external link.
We also present the number of repeated external links that are ever posted in accepted answers.
We observe that for the repeated external links that are uniformly posted different Stack Overflow users, 73.9\% of them are ever posted in accepted answers, i.e., can provide values to Stack Overflow askers.
However, for the repeated external links that only shared by a single Stack Overflow users in answers, 57.3\% of them are ever posted in accepted answers.
This proportion is 0.78 times less than the proportion of repeated external links that are uniformly posted by different Stack Overflow users.
The overall quality of the repeated external links that are shared by the same users is lower than other repeated external links.
This shows that the repeated external links that are shared by the same users contribute less to the crowdsourced knowledge on Stack Overflow.
We suggest that Stack Overflow should pay attention to the repeated external links from a single Stack Overflow user.
For example, Stack Overflow could mark the answers with the external links that are repeatedly shared by a limited number of users to notify viewers.
Viewers are encourage to browse more answers when encounter the answers with the external links that are repeatedly shared by a limited number of users.
To further examine different types of link sharing behavior, we also conduct a qualitative study (see next subsection for details) to understand why users share links.

\begin{figure}[!htb]
  \centering
  \includegraphics[width = \linewidth]{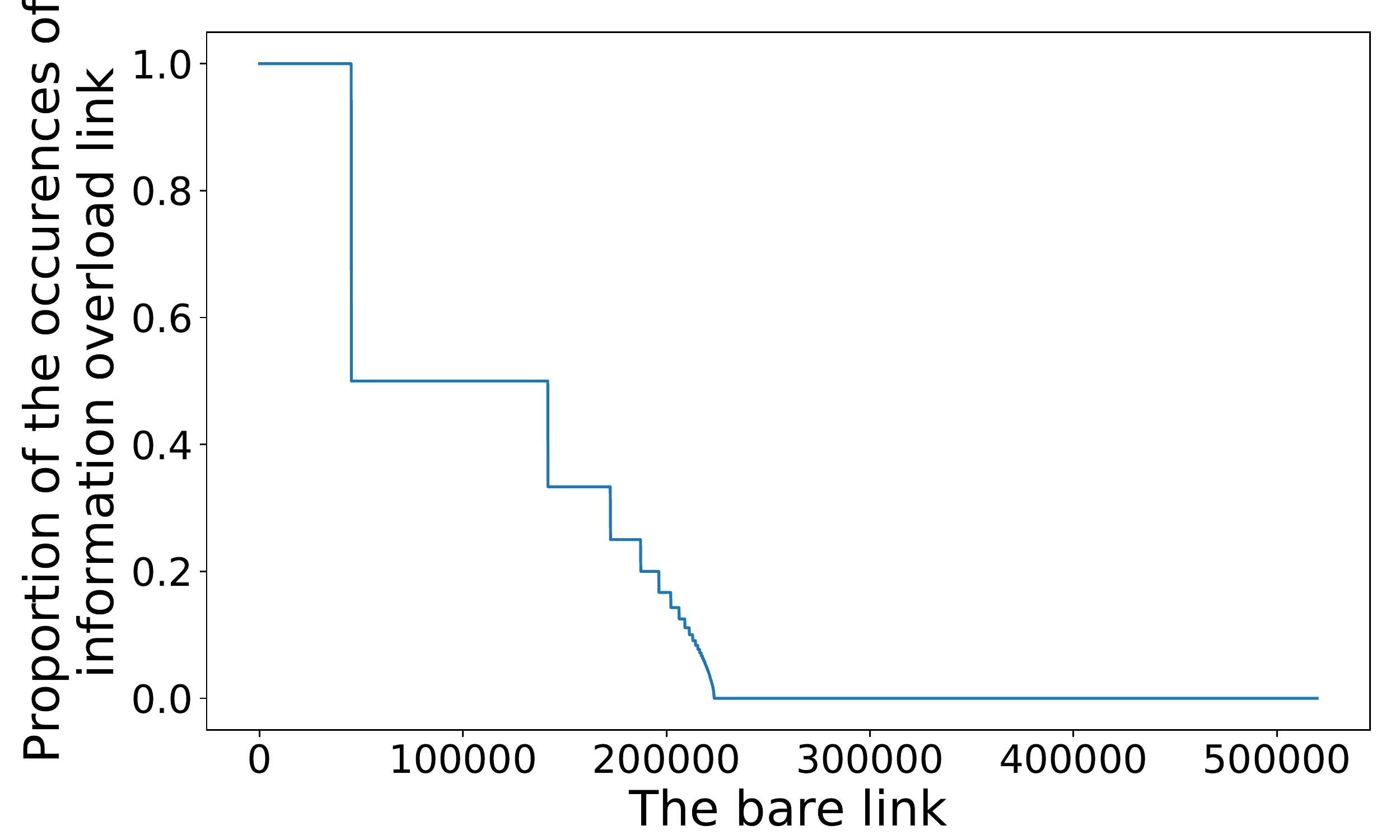}
  \caption{The proportion of the occurrences of the information overloading external link$_{no\_anchor}$.
  In Stack Overflow answers, 41.8\% of the links with anchor part are also shared without anchor part.
  }\label{figure_overload}
\end{figure}

In Stack Overflow answers, \textbf{42.7\% of the external links$_{no\_anchor}$ are shared without anchor part.
The content of webpages shared by external links with no anchor part is overloaded.}
Figure \ref{figure_overload} shows the distribution of the proportions of the occurrences of the information overloading external links$_{no\_anchor}$ among the occurrences of all forms of the external links$_{no\_anchor}$.
On Stack Overflow, 1,590,727 external links$_{anchor}$ locate the resources hosted in 700,379 external links$_{no\_anchor}$.
Among 6,116,153 external links in answers, 704,621 of them are different forms of 524,921 external links$_{no\_anchor}$.
More specifically, 300,920 external links$_{no\_anchor}$ are shared in the form of 480,620 external links$_{anchor}$, and 224,001 external links$_{no\_anchor}$ are shared without showing the specific resources in Stack Overflow.
144,219 (i.e., 64.4\%) of the external links$_{no\_anchor}$ are repeatedly shared in Stack Overflow answers.
12.5\% of the repeated external links in answers are sharing the information overloading links that contain too much irrelevant information.
When users browse the content of the external links$_{no\_anchor}$, it would be difficult for them to retrieve the most relevant part from the webpages.
We suggest that Stack Overflow could design a tool to automatically retrieve the most related content of the links on Stack Overflow.
  
\subsubsection{Qualitative Results}
\textbf{75\% of the repeated external links are shared by different users with the same knowledge to answer different questions.}
For example, answer with id 13733192 provides explanations on whether View.removeAllViews() releases the memory that was used by the child views using below knowledge:

\vspace{0.1cm} \hangindent 1.5em\textit{   A great talk on this can be found on YouTube from Google IO 2011:\\
\underline{http://www.youtube.com/watch?v=\_CruQY55HOk}\\
This talk introduces using the Eclipse MAT (Memory Analysis Tool) to troubleshoot OutOfMemory exceptions.
I suggest it because you mention you're reviewing code and the best place to start looking for an OutOfMemory error is ... "what is taking up my memory?" 
}\vspace{0.1cm}

\noindent\normalsize Answer 15790222 provides solution to destroy old fragments in FragmentStatePagerAdapter using below knowledge:

\vspace{0.1cm} \hangindent 1.5em\textit{   In case you don't know how to analyze memory heap, here is a good \underline{video}.
I can't count how many times it helped me identifying and getting rid of memory leaks in my apps.
}\vspace{0.1cm}

\noindent\normalsize In both answers, a YouTube video\footnote{http://www.youtube.com/wa-tch?v=\_CruQY55HOk} is shared to introduce the memory management in Android.
This external link is repeatedly shared for 98 times in total in Stack Overflow answers, indicating the prevalence of this knowledge.
Different Stack Overflow questions could be answered with the same knowledge that has been posted somewhere else.
Previously, users only can vote on posts and comments.
However, if a Stack Overflow answer cannot be qualified to answer a certain Stack Overflow question, the knowledge hosted in a certain paragraph or sentence in the answer might be used in other questions in the future.
When Stack Overflow users edit the post, Stack Overflow could advise related knowledge based on the relatedness and the prevalence.
One possible approach is that Stack Overflow could design a mechanism where users could vote up or down on a lower granularity, e.g., each paragraph rather than the question or answer.

\textbf{15\% of the sampled post pairs are repeatedly shared by the same Stack Overflow users for promotion purposes.}
For example, a YouTube video\footnote{https://www.youtube.com/playlist?list=PL284C9FF2488-BC6D1} is shared to illustrate the basic use of the GNU Parallel.
This external link is repeatedly shared for 51 times on Stack Overflow, and 50 of the occurrences of this external link is shared by one Stack Overflow user, Ole Tange.
Ole Tange is the author of the video and the author of the GNU Parallel.
The author of the link continually posts the link to promote his work.
One possible reason is that the users try to promote the technology pointed by the link related to the search engine optimization concerns.
If a link is frequently shared by other websites, the search engine would treat the link is valuable and prioritize the rank of the link.
However, only one other user who shares this external link, indicating the value of this external link is not well recognized by other users.
This practice would lead to other useful links hidden within a large number of external links that are repeatedly shared by the same Stack Overflow users.
This makes it difficult for viewers to discover other more relevant links.
We suggest that Stack Overflow could design a mechanism to identify the repeated sharing of external links that are designed for promotion purposes.

\textbf{10\% of the occurrences of the external links are the information overloading links.}
We observe that the shared resources can provide comprehensive information, and much of the information is unnecessary to the concerns of the question, i.e., can be used to answer other questions.
For example, \lstinline{http://git-scm.com/docs/git-rebase} is the link that points to the official documentation webpage that describes the details on the git rebase command.
Such an external link is repeatedly shared in 87 distinct tags, such as git-rebase, branching-and-merging, git-interactive-rebase, and git-filter-branch.
Most of them cover a wide range of knowledge related to the git rebase command.
However, in addition to the description of the command, configurations, and options of Git Rebase, the information of various usage patterns of the command, are also on the same page, e.g., the merge strategies, how to use in interactive mode, and how to recover from upstream rebase.
When viewers browse the posts and click the link that points to the resources that contain too much information, we believe it could be time-consuming for users to search for the relevant information from a lengthy webpage.
In the future, researchers could design an automated tool to retrieve the most relevant information from the content of the linked resources.

\rqbox{Answers with more external links are associated with more revisions, and the use of repeated external links in Stack Overflow answers leads to users repeatedly maintain the same external links in different Stack Overflow answers.
12.3\% of the repeated external links are repeatedly shared by a single Stack Overflow user, indicating that the value of such repeated external links is recognized by a limited number of Stack Overflow users, and can be shared for promotion purposes.
42.7\% of the external links with the anchor part are also shared without anchor part.
The content of webpages shared by external links with no anchor part is often overloaded.
By a qualitative analysis, we observe that 75\% of the repeated external links are shared by different users with the same knowledge to answer different questions.
}

\section{Discussion}\label{s_discuss}
In this section, we discuss the implications of our findings for Stack Overflow moderators and researchers.

\subsection{Implication for Stack Overflow Website Moderators}

\textbf{We encourage Stack Overflow to centrally manage the repeatedly shared knowledge}.
In Section \ref{s_observeing_post}, we observe that different users repeatedly share the same knowledge in the form of repeated external links to answer different questions.
As the knowledge evolves, the content of the repeated external links can be updated, and the knowledge that is hosted within the external links can be obsolete.
For example, \lstinline{http://java.sun.com/j2se/1.4.2/docs/ap}-\lstinline{i/java/nio/MappedByteBuffer.html} is an invalid link and is repeatedly shared for 8 times in history.
2 posts repaired this invalid links by updating this invalid link with \lstinline{http://docs.oracle.com/javase/8/docs/api/java/nio/MappedByteB}-\lstinline{uffer.html}.
In contrast, the rest 6 posts did not maintain the posts with this repeated external link.
This practice increases the maintenance effort of the posts that contain repeated external links, as well as the overall maintenance effort of Stack Overflow.
Stack Overflow can centrally manage and maintain the repeatedly shared knowledge on Stack Overflow.
For example, Stack Overflow could archive the content of the external links and host the archives as internal links.
Users can share the internal links that are maintained by Stack Overflow.
If the question itself is not obsolete (i.e., the question is not related to asking the out of date knowledge), once the external links is updated, the posts with the external links are expected to be updated.

\textbf{We encourage Stack Overflow to notify viewers of the posts with the external links that are repeatedly shared by a limited number of users.}
In Section \ref{s_observeing_post}, we observe that external links can be repeatedly shared by the same Stack Overflow users for promotion purposes.
The value of the external links that are repeatedly shared by the same user is not well recognized by other users.
This practice would lead to other useful links hidden within a large number of external links that are repeatedly shared by the same Stack Overflow users.
This makes it difficult for viewers to discover other more relevant links.
We encourage Stack Overflow to mark the posts with the links that are repeatedly shared by a limited number of users to notify viewers to browse more posts.

\textbf{An automated tool to retrieve the most relevant information from the content of the linked resources is needed.}
In Section \ref{s_observeing_post}, we observe that the shared resources can provide comprehensive information.
Users have to scroll the lengthy resources to retrieve the most related information.
We believe this practice is time-consuming.
Moreover, users are confronted with the risk of failing to locate the most relevant information and the search function provided by the web browsers would be failed.
Therefore, an automated tool is recommended to retrieve the most relevant information from the lengthy content of the linked resource.

\textbf{We encourage Stack Overflow to replace the external links that points to URL shortener websites with the target links}.
In Section \ref{s_observeing_repeat_type}, we observe that the reliability of the links that points to URL shortener websites depends on the accessibility of the target resources and the stability of the URL shortener service provider.
We encourage Stack Overflow to detect the target links of the external links that point to URL shortener websites, and replace the external links that point to URL shortener websites with the target links in full length.

\textbf{We encourage Stack Overflow to proactively notify askers of the existing issues in bug tracking systems.}
In Section \ref{s_observeing_repeat_type}, we observe that Stack Overflow askers may not be aware of the bugs that have already been identified in bug tracker websites before posting a question.
They acknowledge the identified bugs by the external links that are associated with the bug tracker websites in Stack Overflow answers.
We suggest the Stack Overflow could leverage the existing issues from bug tracking systems.
Stack Overflow can display the summary of bugs when a user is in the progress of writing a question that can be related to the bugs in bug tracker websites.

\subsection{Implications for Future Researchers}

\textbf{Researchers could design a tool to help Stack Overflow users better search on search engine}.
In Section \ref{s_observeing_repeat_type}, we observe that Stack Overflow users may not know how to search properly.
Users make a nontrivial effort in searching and still end up asking a question.
In contrast, an expert can solve the question with a better search query.
However, a slight change to the query can lead to a big difference in the search result.
This leads to the case that it is difficult for users to obtain useful search results.
Though previous research investigated the general query reformulation problems in the information retrieval community, there is no research that has investigated how to help developers better search on search engine \citep{dang_queryreformulation}
We suggest that researchers should investigate how to help Stack Overflow users better search on the search engine.

\textbf{Researchers should leverage the target resources linked by external links on Stack Overflow when mining Stack Overflow}.
In Section \ref{s_observeing_repeat_type}, we observe that the external resources can extend the crowdsourced knowledge on Stack Overflow.
However, previous researchers only mined the resources posted within Stack Overflow, ignoring the external resources hosted on external websites.
For example, previous researchers mine the code snippets that are wrapped by the code fragments in the posts bodies when they inspect the code on Stack Overflow \citep{an2017stack,Chaiyong_2018_toxic,Wu2019}.
However, according to their explanations of their dataset, they do not leverage the code resources that are hosted in the external links that point to the code resources.
This practice may lead to bias in their research findings.
We suggest that future researchers should leverage external code resources when they investigate the code that is shared on Stack Overflow.

\section{Threats to Validity}\label{threats}
In this Section, we consider the threats to the validity of our results.

\vspace{0.1cm}\noindent\textbf{Threats to internal validity} concern the factors that could have influenced our results.
We heavily depend on several manual processes in RQs.
Like any human activity, our manual labeling process is subject to personal bias and subjectivity.
To reduce personal bias in the manual labeling process, each labeling object was labeled by two of the authors and discrepancies were discussed until a consensus was reached.
We also showed that the level of inter-rater agreement of the qualitative studies is high (i.e., the values of Cohen's kappa ranged from 0.69 to 0.86).

We characterize the links that are directly presented in the text in Stack Overflow posts and comments.
Though different links can be redirected to the same webpage, we do not consider them as the same link.
This is because that 1) the URLs (i.e., Uniform Resource Locator) of these links are different, indicating that the purpose of sharing these links by users is to share different resources.
Besides, 2) different implementations of URL redirection can cause different result.
Some implementations can make the browser display the URL of the raw links and not the URL of the redirected links in the URL bar; while others may not be.
Currently, there is no technique to detect all cases whether two identical webpages with different URLs are caused by redirected links.
Therefore, we analyze the links that are currently directly presented in the text of Stack Overflow posts and comments.
This could make us observe fewer occurrences of the repeated external links and a fewer number of repeated external links.
In the future, we plan to investigate the repeated shared external webpages by comparing the content of the shared webpages.

%


\vspace{0.1cm}\noindent\textbf{Threats to external validity} concern the generalization of our findings. Our study is conducted to investigate the external links that are directly presented in the text on Stack Overflow. That said, our findings may not be generalized to internal links on Stack Overflow and external links in other Q\&A sites. In the future, we plan to analyze external links in other Q\&A systems.

\section{Conclusion}\label{conclusion}

In this paper, we investigate the repeated external links on Stack Overflow.
We observe that 82.5\% of the link sharing activities (i.e., sharing links in any question, answer, or comment) on Stack Overflow share external resources.
57.0\% of the occurrences of the external links are the repeated external links.
Answers with higher scores have a higher proportion of repeated external links.
External links to text resources (i.e., textual resources that are hosted in documentation websites, tutorial websites, etc.) are the most repeatedly shared.
We also observe that answers with more external links are associated with more revisions.
When the external links are repeatedly shared in different Stack Overflow answers, it is challenging for different users to repeatedly maintain the same external link, thus increasing the overall maintenance efforts.
12.3\% of the repeated external links are repeatedly shared by a single Stack Overflow user, indicating that the value of such repeated external links is recognized by a limited number of Stack Overflow users.
42.7\% of the external links with anchor part are also shared without anchor part.
The content of webpages shared by external links with no anchor part is often highly overloaded.

In the future, we plan to perform a further investigation into how users use one specific type of external link on Stack Overflow, e.g., documentation links, or image links, to disseminate knowledge.

\balance

\bibliographystyle{spbasic}
\bibliography{myreference}

\end{document}